\newif\ifpreprint
\preprinttrue
\ifpreprint
    \documentclass{article}
\else
    \documentclass{hch}
\fi
\usepackage{graphicx}
\usepackage{cite}
\newcommand\dst{dS${}_3$}
\newcommand\dsd{dS${}_d$}
\newcommand\scri{{\cal{I}}}
\begin{document}
\ifpreprint
    \title{Les Houches Lectures on de Sitter Space\footnote{Based
    on lectures by A. S. at the LXXVI
    Les Houches school ``Unity from Duality:  Gravity,
    Gauge Theory and Strings'', August 2001.}}
    \author{Marcus Spradlin${}^{1,2}$, Andrew Strominger${}^1$ and
    Anastasia Volovich${}^1$\footnote{On leave from the L. D. Landau
    Institute for Theoretical Physics, Moscow, Russia.}}
    \date{~}
\else
    \title{De Sitter Space}
    \runningtitle{M. Spradlin, A. Strominger and
    A. Volovich: De Sitter Space}
    \author{Andrew Strominger}
    \authorsup{
	Marcus Spradlin\inst[Harvard University,
	Cambridge MA, USA]{1}${}^,$\inst[Princeton University,
	Princeton NJ, USA]{2},
        Andrew Strominger\inst{1},
        Anastasia Volovich\inst{1}${}^,$\inst[L.D. Landau
        Institute for Theoretical Physics, Moscow, Russia]{3}}
    \address{Department of Physics\\
        Harvard University\\
        Cambridge, MA 02138}
    \thanks{We would like to thank C. Bachas, A. Bilal, F. David, M. Douglas,
    and N. Nekrasov for organizing a very pleasant and productive
    summer school and for arranging financial support.
    This work was supported in part by DOE grant
    DE-FG02-91ER40655.  M.S. is also supported by
    DOE grant DE-FG02-91ER40671, and
    A.V. is also supported by INTAS-OPEN-97-1312.
    We are grateful to R. Bousso and A. Maloney for useful discussions, and
    to C. Herzog and L. McAllister
    for comments on the manuscript.}
\fi
\bigskip
\maketitle
\ifpreprint
   \bigskip
   \bigskip
   \bigskip
   \begin{picture}(0,0)(0,0)
   \put(300,205){PUPT-2015}
   \put(300,220){hep-th/0110007}
   \end{picture}

    \thispagestyle{empty}
    \setcounter{page}{0}
    \smallskip
    {\centerline{\it ${}^{1}$~Department of Physics}}
    {\centerline{\it Harvard University}}
    {\centerline{\it Cambridge, MA 02138}}
    \vskip 0.25in
    {\centerline{\it ${}^{2}$~Department of Physics}}
    {\centerline{\it Princeton University}}
    {\centerline{\it Princeton, NJ 08544}}
    \vskip 0.5in
\fi
\bigskip
\begin{abstract}
These lectures present an elementary discussion of some
background material relevant to the problem of
de Sitter quantum gravity.
The first two lectures discuss the classical geometry of de Sitter space and
properties of quantum field theory on de Sitter space, especially
the temperature and entropy of de Sitter space. The final lecture contains 
a pedagogical discussion of the 
appearance of the conformal group as 
an asymptotic symmetry group, which is central to the dS/CFT
correspondence. A (previously lacking) derivation of 
asymptotically de Sitter boundary
conditions is also given. 
\end{abstract}

\ifpreprint
\newpage
\tableofcontents
\fi
\section{Introduction}

We begin these lectures with one of our favorite equations
\begin{equation}
S = {A \over 4 G}.
\label{bh}
\end{equation}
This is
the Bekenstein-Hawking area-entropy law, which says that the
entropy $S$ associated with an event horizon is its area $A$ divided
by $4 G$, where $G$ is Newton's constant \cite{Bekenstein:1973ur,
Hawking:1975sw}.
This is a
macroscopic formula. It should be viewed in the same
light as the macroscopic thermodynamic formulae that were
first studied in the
18th and 19th centuries.
It describes
how properties of event horizons in general relativity change as
their parameters are varied. This  behavior can
be succinctly summarized by ascribing to them an entropy given by~(\ref{bh}).

One of the surprising
and impressive features of this formula is its universality.
It applies to all kinds of black holes with all kinds of charges,
shapes and  rotation, as well as
to black strings and to all of the strange new
objects we've found in string theory.
It also applies to cosmological
horizons, like the event horizon in de Sitter space \cite{Gibbons:1977mu}.

After Boltzmann's work we tend to think
of entropy in microscopic statistical terms as
something which counts the number of microstates of a system.
Such an interpretation for the entropy~(\ref{bh})  was not given at the time
that the law  was discovered in the early 70s.
A complete understanding
of this law, and in particular of
the statistical origin of this law, is undoubtedly one of the main
keys to understanding what quantum gravity is and what the new notions
are that replace space and time in quantum gravity.

There has been some definite but still limited progress
in understanding the microscopic origin of (\ref{bh}) in
very special cases of black holes which can be embedded into
string theory \cite{Strominger:1996sh}.
That little piece of~(\ref{bh}) that we have
managed to understand has
led to all kinds of interesting insights,
ultimately culminating
in the AdS/CFT
correspondence \cite{Maldacena:1998re}.
Nevertheless the progress towards a complete understanding of~(\ref{bh})
is still very limited, because we only understand
special kinds of black holes---among which Schwarzschild black holes are
not included---and
we certainly don't understand much  about cosmological event horizons,
such as the horizon in de Sitter space.

In some ways cosmological horizons are much more
puzzling than black hole horizons because in the black hole case
one may expect that the black hole
is a localized object with some quantum microstates.
Then if you could provide the correct description of that
localized object, you would be able to count those microstates
and compare your result to the Bekenstein-Hawking formula
and see that they agree.  In
some stringy cases this agreement has been achieved.
On the other hand in de Sitter space the event horizon
is observer dependent, and it is difficult even to see where
the quantum microstates that we would like to count are supposed to be.

Why has there been significant progress in understanding black
hole entropy, but almost no progress in understanding
the entropy of de Sitter space?
One reason is that one of the principal tools we've used
for understanding black
hole entropy is supersymmetry.  Black holes can be supersymmetric, and
indeed the first black holes whose entropy was counted microscopically
were supersymmetric.  Since then we've managed to creep away from
the supersymmetric limit a little bit, but not very far, and
certainly we never managed to get all the way to Schwarzschild black holes.
So supersymmetry is a crutch that we will need to throw away before
we can do anything about de Sitter space.
Indeed there is a very simple observation \cite{Pilch:1985aw}
that de Sitter space is inconsistent
with supersymmetry in the sense that there
is no supergroup that includes
the isometries of de Sitter space and
has unitary representations.\footnote{See, however \cite{Hull:1998vg}.}
A second, related, obstacle to progress in understanding
de Sitter space is that so far we have not been able to
embed it in a fully satisfactory manner into string theory.

While the importance of understanding de Sitter quantum gravity has been
evident for decades,  it has recently been receiving more
attention\cite{Tsamis:1996qm,Maldacena:1998ih,
Hull:1998vg,
dscft,
Witten:2001kn,
wit,
Park:1998qk,
Banados:1999tb,
Kim:1999zs,
Lin:1999gf,
Bousso:1999cb,
Hull:2000mt,
Hawking:2001da,
Banks:2000fe,
Bousso:2001md,
Volovich:2001rt,
Banks:2001yp,
Balasubramanian:2001rb,
Deser:2001us,
Deser:2001xr,
Li:2001ky,
Nojiri:2001mf,
Silverstein:2001xn,
Klemm:2001ea,
Chamblin:2001jj,
Bousso:2000nf,
Gao:2001sr,
Bros:2001yw,
Nojiri:2001qq,
Halyo:2001px,
Sachs:2001qb,
Tolley:2001gg,
Shiromizu:2001bg,
Dolan:2001ih,
Kallosh:2001tm,
Hull:2001ii}.
One reason for this is the recent astronomical observations which indicate
that the cosmological constant in our universe is positive
\cite{Schmidt:1998ys,Riess:1998cb,
Perlmutter:1999np,Perlmutter:2000rr}.
A second reason is
that
recent progress in string theory
and black holes provides new tools and
suggests potentially fruitful new angles. So perhaps de Sitter
quantum gravity is a nut ready to be cracked.
These lectures are mostly an elementary discussion
of the background material relevant to the problem of
de Sitter quantum gravity. The classical geometry of
de Sitter space is described in section \ref{secclassical}.
Scalar quantum field theory
in a fixed de Sitter background is in section \ref{secqft}. Finally,
in section \ref{secqg} we turn to some recent work on de Sitter quantum 
gravity. A pedagogical derivation is given of the appearance of the 
the two dimensional conformal group in three dimensional de Sitter space,
which leads to the dS/CFT correspondence \cite{dscft}.
This section also contains a derivation, missing in previous treatments, of
the asymptotically de Sitter  boundary conditions on the metric. The
appendix 
contains a calculation of the asymptotic form of the Brown-York stress
tensor. 

\section{Classical Geometry of De Sitter Space}
\label{secclassical}

\begin{figure}[hbtp]
    \centering
    \includegraphics[width=1.5truein]{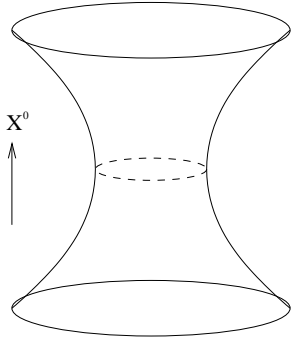}
    \caption[]{Hyperboloid illustrating de Sitter space.
    The dotted line represents an extremal volume $S^{d-1}$.}
    \label{hyperboloid}
\end{figure}

The $d$-dimensional de Sitter space \dsd\ may be
realized as the hypersurface described by the equation
\begin{equation}
-X_0^2+X_1^2+\cdots+X_d^2=\ell^2
\label{hyper}
\end{equation}
in flat $d{+}1$-dimensional Minkowski space ${\cal M}^{d,1}$, where
$\ell$ is a parameter with units of length called the de Sitter
radius.
This hypersurface in flat Minkowski space
is a hyperboloid, as shown in figure \ref{hyperboloid}.

The de Sitter metric is the induced metric from the standard flat
metric on ${\cal{M}}^{d,1}$. The embedding~(\ref{hyper}) is a nice
way of describing de Sitter space because the O$(d,1)$ symmetry,
which is the isometry group of \dsd, is manifest. Furthermore one
can show that \dsd\ is an Einstein manifold with positive scalar
curvature, and the Einstein tensor satisfies
\begin{equation}
G_{ab}+\Lambda g_{ab} = 0,
\label{ein}
\end{equation}
where
\begin{equation}
\Lambda= {(d-2)(d-1) \over 2 \ell^2}
\label{lambda}
\end{equation}
is the cosmological constant.
Henceforth we will set $\ell = 1$.

\subsection{Coordinate Systems and Penrose Diagram}

We will now discuss a number of  coordinate systems on \dsd\ which
give different insights into the structure of \dsd. We will
frequently make use of coordinates on the sphere $S^{d-1}$, which
is conveniently parametrized by setting
\begin{eqnarray}
\omega^1 &=& \cos \theta_1,\cr
\omega^2 &=& \sin \theta_1 \cos \theta_2,\cr
&\vdots&\cr
\omega^{d-1} &=& \sin \theta_1 \cdots \sin \theta_{d-2} \cos \theta_{d-1},\cr
\omega^{d} &=& \sin \theta_1 \cdots \sin \theta_{d-2} \sin \theta_{d-1},
\label{omegas}
\end{eqnarray}
where $0 \le \theta_i < \pi$ for  $1 \le i < d-1$, but $0 \le \theta_{d-1} < 2
\pi$.
Then it is clear that $\sum_{i=1}^{d} (\omega^i)^2 = 1$, and the metric on
$S^{d-1}$ is
\begin{equation}
d \Omega_{d-1}^2 = \sum_{i=1}^d (d\omega^i)^2
= d \theta_1^2 + \sin^2 \theta_1 d \theta_2^2 + \cdots +
\sin^2 \theta_1 \cdots \sin^2 \theta_{d-2}  d \theta_{d-1}^2.
\end{equation}

\vskip .15in
\noindent
{\bf a. Global coordinates $(\tau, \theta_i)$.}
This coordinate system is obtained by setting
\begin{eqnarray}
X^0 &=& \sinh \tau,\cr
X^i &=& \omega^i \cosh \tau, ~~~~~ i=1,\ldots,d,
\end{eqnarray}
\label{ccv}
where $-\infty < \tau <\infty$ and the $\omega^i$ are as
in~(\ref{omegas}).
It is not hard to check that these
satisfy~(\ref{hyper}) for any point $(\tau, \omega_i)$.

 From the flat metric on ${\cal{M}}^{d,1}$
\begin{equation}
ds^2=-dX_0^2+dX_1^2+\cdots+dX_d^2,
\label{metricf}
\end{equation}
plugging in (\ref{ccv}) 
we obtain the induced metric
on \dsd,
\begin{equation}
ds^2=-d\tau^2+(\cosh^2 \tau) d\Omega_{d-1}^2.
\label{metricg}
\end{equation}
In these coordinates \dsd\ looks like a $d{-}1$-sphere which starts
out infinitely large at $\tau = -\infty$, then shrinks to a minimal
finite size at $\tau = 0$, then grows
again to infinite size as $\tau \to +\infty$.

\vskip .15in
\noindent
{\bf b. Conformal coordinates $(T,\theta_i)$.}
These coordinates are related to the global coordinates by
\begin{equation}
\cosh \tau={1 \over \cos T},
\label{ttau}
\end{equation}
so that we have $-\pi/2 < T< \pi/2.$
The metric in these coordinates takes the form
\begin{equation}
ds^2={1 \over \cos^2 T}(-dT^2+d\Omega_{d-1}^2).
\label{metricc}
\end{equation}
This is a particularly useful coordinate system
because it enables us to understand the causal structure
of de Sitter space.
If a geodesic is null with respect to the metric~(\ref{metricc}),
then it is also null with respect to the conformally related metric
\begin{equation}
d{\tilde s}^2=(\cos^2 T) ds^2=-dT^2+d\Omega_{d-1}^2.
\label{metrict}
\end{equation}
So from the point of view of analyzing what null geodesics do
in \dsd\
we are free to work with the metric~(\ref{metrict}), which looks simpler
than~(\ref{metricc}).

\begin{figure}[hbtp]
    \centering
    \includegraphics[width=1.55truein]{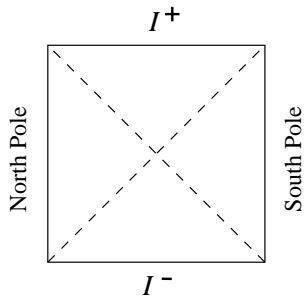}
    \caption[]{Penrose diagram for \dsd. The north and south poles
are timelike lines, while every point in the interior represents
an
    $S^{d-2}$.
A horizontal slice is an $S^{d-1}$. The dashed lines are the past and future
    horizons of an observer at the south pole.  The conformal
    time coordinate $T$ runs from $-\pi/2$ at $\scri^-$ to
    $+\pi/2$ at $\scri^+$.}
    \label{penrose}
\end{figure}

The Penrose diagram \ref{penrose}
contains all the information about the causal structure of
\dsd\
although distances are highly distorted.
In this diagram each point is actually an $S^{d-2}$
except for points on the left or right
sides, which lie on the north or south pole respectively.
Light rays travel at $45^\circ$ angles in this diagram, while
timelike surfaces are more vertical than horizontal and spacelike
surfaces are more horizontal than vertical.

The surfaces marked $\scri^-$, $\scri^+$ are called past and future null
infinity.  They are the surfaces where all null geodesics originate
and terminate.
Note that a light ray which starts at the north pole at $\scri^-$
will exactly reach the south pole by the time it reaches $\scri^+$
infinitely far in the future.

\begin{figure}[hbtp]
    \centering
    \includegraphics[width=3.8truein]{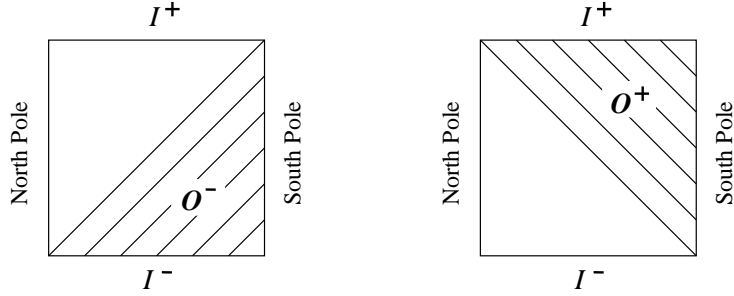}
    \caption[]{These diagrams show the regions $\cal{O}^-$
    and $\cal{O}^+$ corresponding respectively to the causal
    past and future of an observer at the south pole.}
    \label{penrosetwo}
\end{figure}

One of the peculiar features of  de Sitter space is that no single
observer can access the entire spacetime.
We often get into trouble in physics
when we try to describe
more than we are allowed to observe---position and momentum in
quantum mechanics, for example. Therefore in attempting to
develop de Sitter quantum gravity we should be aware of
what can and cannot be observed.
A classical  observer sitting on the south pole
will never be able to observe anything past the diagonal
line stretching from the north pole at $\scri^-$ to the south
pole at $\scri^+$.
This region is marked as ${\cal{O}}^-$ in figure \ref{penrosetwo}.
This is qualitatively different from Minkowski space, for example,
where a timelike observer will eventually have the entire history of the
universe in his/her  past light cone.

Also shown in figure \ref{penrosetwo} is the region ${\cal{O}}^+$, which
is the only part of de Sitter space that an observer on the south
pole will ever be able to send a message to.
The intersection ${\cal O}^+ \cap {\cal O}^-$
is called the (southern) causal diamond.  It is this region that is
fully accessible to the observer on the south
pole.  For example if she/he wishes to know the weather anywhere in the
southern diamond, a query can be sent to the appropriately located weather
station and the response received before $\scri^+$ is reached.
This is not possible in the lower diamond of ${\cal{O}}^-$,
to which a query can never be sent, or the upper diamond of ${\cal{O}}^+$,
from
which a response cannot be received.
The
northern diamond on the left of \ref{penrosetwo} is completely
inaccessible to
an observer on the south pole.

\begin{figure}[hbtp]
    \centering
    \includegraphics[width=1.85truein]{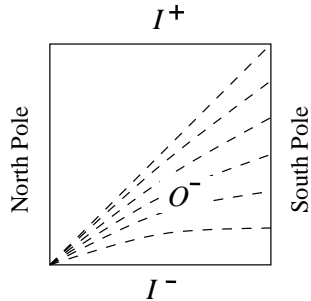}
    \caption[]{The dashed lines are slices of constant
    $t$ in planar coordinates.  Note that each slice is
    an infinite flat $d{-}1$-dimensional plane which extends
    all the way down to $\scri^-$.}
    \label{planar}
\end{figure}

\vskip .15in
\noindent
{\bf c. Planar coordinates $(t,x^i)$, $i=1,\ldots,d-1$.}
To define this coordinate system we take
\begin{eqnarray}
X^0 &=& \sinh t - \frac{1}{2} x_i x^i e^{-t},\cr
X^i &=&  x^i e^{-t}, ~~~~~ i=1,\ldots,d-1,\cr
X^d &=& \cosh t - \frac{1}{2} x_i x^i e^{-t}.
\end{eqnarray}
The metric then takes the form
\begin{equation}
ds^2=-dt^2+e^{-2t} dx_i dx^i.
\label{metricp}
\end{equation}
These coordinates do not cover all of de Sitter space, but only
the region ${\cal O}^-$ and are therefore
appropriate for an observer on the south pole.
The slices of constant $t$ are illustrated in
figure \ref{planar}.

The surfaces of constant $t$ are spatial sections of de Sitter
space which are infinite volume $d{-}1$-planes with the flat metric.
 From the diagram it is clear that every surface of constant $t$
intersects $\scri^-$ at the north pole.
It may seem puzzling---and is certainly one of the salient features of
de Sitter space---that a spatial plane can make it to
the infinite past.
This happens because $\scri^-$
is very large, and you can get there along a spatial trajectory
from anywhere in ${\cal O}^-.$
In these coordinates the time $t$
is not a Killing vector, and the only manifest symmetries
are translations and rotations of the $x^i$ coordinates.

\vskip .15in
\noindent
{\bf d. Static coordinates $(t,r,\theta_a)$, $a=1,\ldots,d-2$.}
The $t$ in these coordinates is not the same as the $t$ in
planar coordinates, but we are running out of letters!  Note
also that for these coordinates and the following ones we
will need a parametrization of $S^{d-2}$, not $S^{d-1}$.
This coordinate system is constructed to have an explicit
timelike Killing symmetry.
If we write
\begin{eqnarray}
X^0 &=& \sqrt{1-r^2} \sinh t,\cr
X^a &=& r \omega^a, ~~~~~  a = 1,\ldots,d-1,\cr
X^d &=& \sqrt{1 - r^2} \cosh t,
\end{eqnarray}
then the metric takes the form
\begin{equation}
ds^2=-(1-r^2)dt^2+\frac{dr^2}{1-r^2} +r^2 d\Omega_{d-2}^2.
\label{metrics}
\end{equation}
In this coordinate system $\partial/\partial t$
is a Killing vector and generates the symmetry
$t \to t + {\rm constant}$.
The horizons are at $r^2 = 1$, and the southern
causal diamond has $0 \le r \le 1$, with the south
pole at $r = 0$.

\begin{figure}[hbtp]
    \centering
    \includegraphics[width=1.7truein]{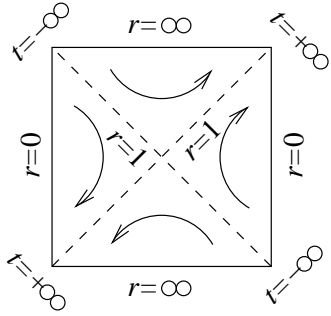}
    \caption[]{This Penrose diagram shows the direction of
    the flow generated by the Killing vector $\partial/\partial t$ in
    static coordinates.  The horizons (dotted lines) are at
    $r^2 =  1$, and the southern causal diamond
    is the region with $0 \le r \le 1$ on the right hand side.  Past and future
    null infinity $\scri^\pm$ are at $r = \infty$.}
    \label{static}
\end{figure}

One of the reasons to want a timelike Killing vector is
so that we can use it to define time evolution, or in
other words to define the Hamiltonian.
But from~(\ref{metrics}) we see that at $r=1$ the
norm of $\partial /\partial t$ vanishes, so that it becomes null.
In figure \ref{static} we illustrate what the Killing vector
field $\partial/\partial t$ is doing when extended to the
various diamonds of the
Penrose diagram.  In the top and bottom diamonds, $\partial/\partial t$
is spacelike,
while in the northern diamond the vector is pointing towards the past!
Thus $\partial/\partial t$ in static coordinates can only be used to
define a sensible time evolution in the southern
diamond of de Sitter space. The absence of a globally timelike Killing
vector in de Sitter space has important implications for the quantum
theory.

\vskip .15in
\noindent
{\bf e. Eddington-Finkelstein coordinates $(x^+,r,\theta_a)$.}
This coordinate system is the de Sitter analog of the (outgoing)
Eddington-Finkelstein coordinates for a Schwarzschild black hole.
Starting from the static coordinates, we define
$x^+$ by the equation
\begin{equation}
dt=dx^++\frac{dr}{1-r^2},
\label{dtt}
\end{equation}
which we can solve to obtain
\begin{equation}
x^+=t+\frac{1}{2} \ln \frac{1+r}{1-r}.
\end{equation}
In these coordinates the metric is
\begin{equation}
ds^2=-(1-r^2) (dx^+)^2-2 dx^+ dr +r^2 d\Omega_{d-2}^2.
\end{equation}
The same symmetries are manifest in this coordinate
system as in the static coordinates since
$\partial/\partial t$ at fixed $r$
is the same as $\partial/\partial x^+$
at fixed $r$. Lines of constant $x^+$ are the null lines connecting
${\cal I}^-$ with the south pole depicted in figure \ref{penrosetwo}.
These coordinates cover the causal past ${\cal{O}}^-$
of an observer at the south pole while still keeping the symmetry
manifest.

We can also define
\begin{equation}
x^-=t-\frac{1}{2}\ln \frac{1+r}{1-r},
\end{equation}
so that the metric takes the form
\begin{equation}
ds^2=-(1-r^2(x^+,x^-)) dx^+ dx^- +r^2 d\Omega_{d-2}^2,
\end{equation}
where $r= \tanh {x^+-x^- \over 2}$.

\vskip .15in
\noindent
{\bf f. Kruskal coordinates $(U,V,\theta_a)$.}
Finally we take
\begin{equation}
x^-=\ln U,~~~x^+=-\ln(-V),
\end{equation}
in which case
\begin{equation}
r=\frac{1+UV}{1-UV}.
\end{equation}
Then the metric takes the form
\begin{equation}
ds^2=\frac{1}{(1-UV)^2}(-4 dU dV +(1+UV)^2 d\Omega_{d-2}^2).
\end{equation}
These coordinates cover all of de Sitter space.
The north and south poles correspond to $UV = -1$, the horizons
correspond to $UV = 0$, and $\scri^\pm$ correspond to $UV = 1$.
The southern diamond is the region with $U>0$ and $V<0$.

{\it Exercise~1.}
Find $X^0, \ldots, X^d$
as functions of $U, V$, and $\theta_a$ for the Kruskal coordinates.

\begin{figure}[hbtp]
        \centering
        \includegraphics[width=1.5truein]{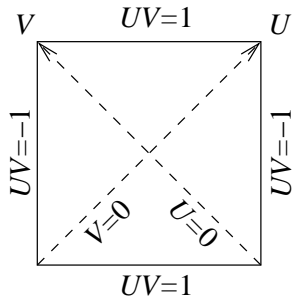}
        \caption[]{The Kruskal coordinate system covers all
    of de Sitter space.  In this Penrose diagram the coordinate
    axes $U=0$ and $V=0$ are the horizons, $UV = -1$ are the
    north and south poles, and $UV = 1$ are $\scri^+$ and
    $\scri^-$.  The arrows denote the directions of increasing
    $U$ and $V$.}
        \label{kruskal}
\end{figure}

\vskip .15in
\noindent
{\bf g. Hyperbolic coordinates $(\overline{\tau}, \psi, \theta_a)$}.
Global coordinates foliate de Sitter space with spheres,
and planar coordinates foliate de Sitter space with
planes.
One can also foliate de Sitter space with spaces of constant
negative curvature by using the hyperbolic coordinates
\begin{eqnarray}
X^0 &=& \sinh \overline{\tau} \cosh\psi,\cr
X^a &=& \omega^a \sinh \overline{\tau} \sinh \psi,\cr
X^d &=& \cosh \overline{\tau},
\end{eqnarray}
in which the metric takes the form
\begin{equation}
ds^2 = - d\overline{\tau}^2 + \sinh^2 \overline{\tau}
(d \psi^2 + \sinh^2 \psi\,d\Omega_{d-2}^2),
\end{equation}
and surfaces of constant $\overline{\tau}$ are
$d{-}1$-dimensional hyperbolic planes.

\subsection{Schwarzschild-de Sitter}

The simplest generalization of the de Sitter space solution is
Schwarzschild-de Sitter, which we abbreviate as SdS.
This solution represents a black hole in de Sitter space.
In $d$ dimensions in static coordinates the
SdS${}_d$
metric takes the form
\begin{equation}
ds^2=-(1-\frac{2m}{r^{d-3}}-r^2)\  dt^2 +
\frac{1}{1-\frac{2m}{r^{d-3}}-r^2}\  dr^2+r^2 d\Omega_{d-2}^2,
\end{equation}
where $m$ is a parameter related to the black hole mass (up to
some $d$-dependent normalization constant).
In general there
are two horizons (recall that
these are  places where the timelike Killing vector
$\partial /\partial t$ becomes null), one of which is the black hole
horizon and the other of which is the de Sitter horizon.
Note that the two horizons approach each other as $m$
is increased, so that there is a maximum size black hole
which can fit inside de Sitter space before the black hole
horizon hits the de Sitter horizon.

One reason to introduce SdS
is that it plays an important
role in the work of Gibbons and Hawking \cite{Gibbons:1977mu}
determining the entropy of pure de Sitter space, which will
be reviewed in subsection \ref{entropy}.
For this purpose it will be convenient to focus on the
three dimensional Schwarzschild-de Sitter solution \cite{Deser:1984dr}
\begin{equation}
ds^2 = - (1 - 8 G E - r^2) dt^2 + {dr^2 \over (1 - 8 G E - r^2)}
+ r^2 d\phi^2,
\label{sdsthree}
\end{equation}
where we have normalized the energy $E$ of the Schwarzschild
black hole appropriately for three dimensions.
In three dimensions there is only one horizon, at $r_H
= \sqrt{1 - 8 G E}$, and as $E$ goes to zero this reduces
to the usual horizon in empty de Sitter space.
The fact that there is only a de Sitter horizon and not a black
hole horizon is not surprising in light of the fact that in
three dimensional flat space there are no black holes.

We can learn a little more about the solution~(\ref{sdsthree})
by looking near $r = 0$, where $ds^2$ behaves like
\begin{equation}
ds^2\sim -r_H^2  dt^2+\frac{dr^2}{r_H^2}+r^2 d\phi^2.
\label{nearzero}
\end{equation}
Now we can rescale the coordinates by defining
\begin{equation}
t' = r_H t, ~~~~~
r' = r/r_H, ~~~~~
\phi' =  r_H \phi.
\end{equation}
In the rescaled coordinates the metric~(\ref{nearzero})
is simply
\begin{equation}
ds^2 = - dt'^2 + dr'^2 + r'^2 d\phi'^2.
\end{equation}
This looks like flat space, but it is not quite flat space because
while $\phi$ was identified modulo $2 \pi$, $\phi'$ is identified
modulo $2 \pi r_H$.
Therefore there is a conical singularity with a positive deficit
angle at the origin.

You may be familiar with the fact that
if you put a point-like mass in flat three dimensional Minkowski
space you would also get a conical deficit angle at the location
of the particle. Hence we recognize~(\ref{sdsthree}) as a point-like mass,
rather than a black hole, at the south pole  of \dst. If the solution is
maximally extended one finds there is also point-like mass of the same size
at the north pole \cite{Deser:1984dr}.

{\it Exercise~2.}
Show that SdS${}_3$ is a global identification of
\dst .

\subsection{Geodesics}

Our last topic in the classical geometry of de Sitter space
is geodesics.
It is clear that if we take two points on the sphere $S^n$ of radius $R$,
then there is only one independent SO$(n{+}1)$-invariant quantity that we can
associate to the two points.  That is the geodesic distance
$D$, or equivalently the angle $\theta$ between them, which are
related by $D = R \theta$.
Let us think of the sphere as being embedded in flat
Euclidean space, with the embedding equation
$\delta_{ij}X^i X^j = R^2$, $i,j = 1,\ldots,n{+}1$.
It is useful to define a quantity $P$ by
$R^2 P(X,X') \equiv \delta_{ij} X^i X^{\prime j}
= R^2 \cos \theta$.

It is a little harder to visualize, but we can do something
similar for \dsd.
There we can define
\begin{equation}
P(X,X') = \eta_{ij} X^i X^{\prime j},
~~~~~ \eta_{ij} = {\rm diag}(-1,1,\ldots,1)
\end{equation}
(recall that we have set the de Sitter radius $\ell $ to one).
For points in a common causal diamond, 
this is related to the geodesic distance $D(X,X')$ between $X$
and $X'$ by
$P = \cos D$.
This quantity $P$ will turn out to be a more convenient invariant
to associate to two points in de Sitter space.
We can easily
write explicit formulas for $P(X,X')$ in the various coordinate
systems discussed above.  For example, in planar
coordinates we have
\begin{equation}
P(t,x^i;t',y^i) =
\cosh(t-t') - {1 \over 2} e^{-t-t'} \delta_{ij} (x^i-y^i)(x^j-y^j).
\label{useless}
\end{equation}
The expression for $P$
is simple in terms of the $X$'s but
can get complicated when written in a particular
coordinate system.

To conclude, we note a few important properties of $P$
for later use.  If $P = 1$, then the geodesic distance is equal
to zero, so the two points $X$ and $X'$ coincide or are
separated by a null geodesic.
We can also consider taking antipodal points
$X' = - X$, in which case $P = - 1$.
In general $P = - 1$ when the antipodal point of $X$ lies on
the light cone of $X'$.
In general, the geodesic separating $X$ and $X'$ is
spacelike for $P < 1$ and timelike for $P > 1$, while
for $P < -1$ the geodesic between $X$ and the antipodal point
of $X'$ is timelike.

\section{Quantum Field Theory on De Sitter Space}
\label{secqft}

Ultimately, a complete understanding of the entropy-area
relation~(\ref{bh}) in de Sitter space will require
an understanding of quantum gravity
on de Sitter space.  In this section we will take
a baby step in that direction by considering a single
free massive scalar field on a fixed background
de Sitter spacetime.
This turns out to be a very rich subject
which has been studied by many authors
\cite{birdav,Gibbons:1977mu,Chernikov:1968zm,
Tagirov:1973vv,Figari:1975km,Rumpf:1978bv,
Abbott:1982ff,
Najmi:1984vn,Ford:1985hs,allen,mottola,Allen:1987tz}.

\subsection{Green Functions and Vacua}
\label{vacua}

Let us consider a scalar field in \dsd\ with
the action
\begin{equation}
S = - \frac{1}{2} \int  d^dx \sqrt{-g} \left[
(\nabla \phi)^2 + m^2 \phi^2 \right].
\end{equation}
Since this is a free field theory, all information is
encoded in the two-point function of $\phi$.
We will study the Wightman function
\begin{equation}
G(X,Y)=\langle 0| \phi(X) \phi(Y) |0 \rangle,
\label{wightmann}
\end{equation}
which obeys
the free field equation
\begin{equation}
(\nabla^2 -m^2) G(X,Y)=0,
\label{freefield}
\end{equation}
where $\nabla^2$ is the Laplacian on \dsd.

There are other two point functions that one can
discuss: retarded, advanced, Feynman, Hadamard and so on,
but these can all be obtained from the Wightman function~(\ref{wightmann}),
for example by taking the real or imaginary part, and/or
by multiplying by a step function in time.

Let us assume that the state
$|0 \rangle$ in~(\ref{wightmann}) is invariant under
the SO$(d,1)$ de Sitter group. Then $G(X,Y)$ will
be de Sitter invariant, and so at generic points  can only depend
on the de Sitter invariant length $P(X,Y)$ between
$X$ and $Y$.\footnote{ $P(X,Y)=P(Y,X)$ is insensitive to the 
time ordering between points, which is SO$(d,1)$ (but not O$(d,1)$)
invariant. Because of this the $i\epsilon$ prescription for $G$, as
discussed below, can not be written as a function of $P$ alone.}
Writing $G(X,Y) = G(P(X,Y)),$
(\ref{freefield}) reduces to a differential
equation in one variable $P$
\begin{equation}
(1 - P^2) \partial_P^2 G - d P \partial_P G - m^2 G = 0.
\label{peqn}
\end{equation}
With the change of variable $z = \frac{1 + P}{2}$
this becomes a hypergeometric equation
\begin{equation}
z(1-z) G'' + (\frac{d}{2} - d z) G' - m^2 G = 0,
\end{equation}
whose solution is
\begin{equation}
G =
c_{m,d}
F(h_+, h_-, \frac{d}{2}, z),
\label{hypgeo}
\end{equation}
where $c_{m,d}$ is a normalization constant to be fixed shortly, and
\begin{equation}
h_\pm = \frac{1}{2}\left[ (d-1) \pm \sqrt{(d-1)^2 - 4 m^2}\right].
\end{equation}
The hypergeometric function~(\ref{hypgeo}) has a singularity
at $z = 1$, or $P = 1$, and a branch cut for $1 < P < \infty$.
The singularity occurs when the
points $X$ and $Y$ are separated by a null geodesic.
At short distances the scalar field is insensitive
to the fact that it is in de Sitter space and the
form of the singularity is precisely the same as that
of the propagator in flat $d$-dimensional Minkowski
space.
We can use this fact to fix
the normalization constant $c_{m,d}$.  Near $z = 1$
the hypergeometric function behaves as
\begin{equation}
F(h_+, h_-, \frac{d}{2}, \frac{1+P}{2}) \sim
\left( {D^2 \over 4} \right)^{1-d/2}
\frac{ \Gamma(\frac{d}{2}) \Gamma({d \over 2} - 1)}
{\Gamma(h_+) \Gamma(h_-)},
\end{equation}
where $D = \cos^{-1} P$ is the geodesic separation between the two points.
Comparing with the usual short-distance singularity
$\frac{\Gamma({d \over 2})}
{2(d-2) \pi^{d/2}} (D^2)^{1-d/2}$
fixes the coefficient to be
\begin{equation}
c_{m,d} = 4^{1 - d/2}
\frac{\Gamma(h_+) \Gamma(h_-)}{ \Gamma(\frac{d}{2}) \Gamma(
\frac{d}{2}-1)} \times \frac{\Gamma(\frac{d}{2})}{ 2(d-2) \pi^{d/2}}
=  \frac{\Gamma(h_+) \Gamma(h_-)}{(4 \pi)^{d/2} \Gamma(\frac{d}{2})}.
\end{equation}
The prescription for going around the singularity in the complex plane
is also
the same as in Minkowski space, namely replacing $X^0-Y^0$ with $X^0-Y^0-
i\epsilon $.
%---namely, we
%replace $z$ by $z - i \epsilon\, {\rm sign}(X^0{-}Y^0)$
%in~(\ref{hypgeo}).

The equation~(\ref{peqn}) clearly has a $P \to -P$ symmetry,
so if $G(P)$ is a solution then $G(-P)$ is also a solution.
The second linearly independent solution to~(\ref{peqn})
is therefore
\begin{equation}
F(h_+,h_-,\frac{d}{2},\frac{1-P}{2}).
\label{greentwo}
\end{equation}
The singularity is now at $P = -1$, which corresponds
to $X$ being null separated from the antipodal point
to $Y$.
This singularity sounds rather unphysical at first,
but we should recall that antipodal points in de Sitter
space are always separated by a horizon.  The
Green function~(\ref{greentwo}) can be thought of
as arising from an image source
behind the horizon, and~(\ref{greentwo}) is nonsingular
everywhere within an observer's horizon. Hence the 
``unphysical'' singularity can not be detected by any experiment. 

De Sitter space therefore has a one parameter
family of de Sitter invariant Green functions $G_\alpha$
corresponding
to a linear combination of the
solutions~(\ref{hypgeo}) and~(\ref{greentwo}).
Corresponding to this one-parameter family
of Green functions is a one-parameter family of
de Sitter invariant vacuum states $|\alpha\rangle$
such that $G_\alpha(X,Y) = \langle \alpha| \phi(X) \phi(Y)|\alpha\rangle$.
These vacua are discussed in detail in \cite{allen, mottola}, but
are usually discarded as somehow ``unphysical''. However,
as we try to understand the quantum theory of de Sitter space
these funny extra vacua will surely turn out
to have some purpose in life.

De Sitter Green functions are often discussed
in the context of analytic continuation to the Euclidean sphere.
If we work in static coordinates and take
$t\to i \tau$, the \dsd\
metric becomes the metric on the sphere $S^d$.
On the sphere there is a unique Green function, which
when analytically continued back to de Sitter space
yields~(\ref{hypgeo}).

Let us say a few more words about the vacuum states.
A vacuum state $|0\rangle$ is defined as usual by
saying that it is annihilated by all annihilation operators
\begin{equation}
a_n|0\rangle=0.
\end{equation}
That is, we write an expansion for the scalar field
in terms of creation and annihilation operators
of the form
\begin{equation}
\phi(X) = \sum_k \left[a_k u_k(X) + a_k^\dagger u_k^*(X)\right],
\label{phifield}
\end{equation}
where $a_k$ and $a_k^\dagger$ satisfy
\begin{equation}
[a_k,a^{\dagger}_l]=\delta_{kl}.
\end{equation}
The
modes $u_k(X)$ satisfy
the wave equation
\begin{equation}
(\nabla^2 -m^2) u_k=0,
\end{equation}
and are normalized with respect to the invariant
Klein-Gordon inner product
\begin{equation}
(u_k,u_l)=-i \int d\Sigma^\mu\ \left( u_k
\stackrel{\leftrightarrow}{\partial}_\mu u_l^*\right)=\delta_{kl},
\end{equation}
where the integral is taken over a complete spherical 
spacelike slice in \dsd\ and
the result is independent of the choice of this slice.

The question is, which modes do we associate with
creation operators in~(\ref{phifield}) and which do we
associate with annihilation operators?
In Minkowski space we take
positive and negative frequency modes,
\begin{equation}
u \sim e^{-iEt} f(x),~~~~~u^* \sim e^{iEt} f^*(x),
\end{equation}
respectively to multiply the annihilation and creation
operators.
But in a general curved spacetime there is no canonical
choice of a time variable with respect to which one can
classify modes as being positive or negative frequency.
If we make a choice of time coordinate, we can get a vacuum
state $|0\rangle$ and then the state $(a^{\dagger})^n
|0\rangle \equiv |n\rangle$
is said to have $n$ particles in it.  But if
we had made some other choice of time coordinate then
we would have a different vacuum $|0'\rangle$, which we could
express as a linear combination of the $|n\rangle$'s.
Hence the question ``How many particles are present?'' is not
well-defined independently of a choice of coordinates.  This
is an important and general feature of quantum field theory
in curved spacetime.

In order to preserve classical  symmetries
of \dsd\ in the quantum theory,
we would like to find a way to divide the modes into
$u$ and $u^*$ that is invariant under SO$(d,1)$.
The resulting vacuum will then be de Sitter
invariant.
It turns out \cite{Chernikov:1968zm,
allen,mottola} that there is a family of such
divisions, and a corresponding family of Green functions
such as $G_\alpha $.

\subsection{Temperature}

In this section we will show that an observer moving
along a timelike
geodesic observes a thermal bath of
particles when the scalar field $\phi$ is in the
vacuum state $|0\rangle$.
Thus we will
conclude that de Sitter space is naturally
associated with a temperature \cite{Figari:1975km},
which we will calculate.

Since the notion of a particle is observer-dependent in
a curved spacetime, we must be careful to give a
coordinate invariant characterization of the temperature.
A good way to achieve this is to consider an observer
equipped with a detector.
The detector will have some internal energy states and
can interact with the scalar field by exchanging
energy, $i.e.$ by emitting or absorbing scalar particles.
The detector could for example be constructed
so that it emits a `bing' whenever its internal
energy state changes. All observers will agree on
whether or not the detector has binged, although they may disagree
on whether the bing was caused by particle emission or absorption.
Such a
detector is called an Unruh detector and
may be modeled by a coupling
of the scalar
field $\phi(x(\tau))$ along the worldline $x(\tau)$
of the
observer to some operator
$m(\tau)$ acting on the internal detector states
\begin{equation}
g \int_{-\infty}^\infty d\tau\ m(\tau) \phi(x(\tau)),
\end{equation}
where $g$ is the strength of the coupling and $\tau$
is the proper time along the observer's worldline.

Let $H$ denote the detector Hamiltonian, with energy
eigenstates $|E_j\rangle$,
\begin{equation}
H|E_j \rangle=E_j |E_j \rangle,
\end{equation}
and let $m_{ij}$ be the matrix elements of the operator
$m(\tau)$ at $\tau = 0$:
\begin{equation}
m_{ij} \equiv\langle E_i | m(0) |E_j \rangle.
\end{equation}
We will calculate the transition amplitude from a state
$|0\rangle |E_i\rangle$ in the tensor product of the scalar field and
detector Hilbert spaces to the state $\langle E_j| \langle \beta |$,
where $\langle \beta|$ is any state of the scalar field.
To first order in perturbation theory for small coupling $g$,
the desired amplitude is
\begin{equation}
g \int_{-\infty}^\infty d\tau\ \langle E_j| \langle
\beta | m(\tau) \phi(x(\tau))
|0\rangle |E_i\rangle.
\end{equation}
Using
\begin{equation}
m(\tau)=e^{iH \tau} m(0)e^{-iH \tau},
\end{equation}
this can be written
as
\begin{equation}
g m_{ji} \int_{-\infty}^\infty d\tau\ e^{i(E_j-E_i) \tau}
 \langle \beta | \phi(x(\tau)) |0\rangle.
\label{amplitude}
\end{equation}
Since we are only interested in the probability for the
detector to make the transition from $E_i$ to $E_j$, we should
square the amplitude~(\ref{amplitude}) and sum
over the final state $|\beta\rangle$ of the scalar field,
which will not be measured.
Using $\sum_\beta |\beta \rangle \langle \beta | = 1$, we find
the probability
\begin{equation}
P(E_i \to E_j) = g^2 |m_{ij}|^2 \int_{-\infty}^\infty d\tau \,d\tau'\
e^{-i(E_j - E_i) (\tau'-\tau)} G(x(\tau'), x(\tau)),
\label{integral}
\end{equation}
where $G(x(\tau'), x(\tau))$ is the Green function~(\ref{wightmann}).
The Green function is a function only of the geodesic
distance $P(x(\tau), x(\tau'))$, and if we consider for simplicity
an observer sitting on the south pole, then $P$ is given in static
coordinates by
$P = \cosh(\tau - \tau')$.  Therefore everything inside
the integral~(\ref{integral}) depends only on $\tau-\tau'$ and
we get an infinite factor from integrating over $\tau+ \tau'$.
We can divide out this factor and discuss the transition 
probability per
unit proper time along the detector worldline, which is then
given by
\begin{equation}
\dot{P}(E_i \to E_j) = g^2 |m_{ij}|^2 \int_{-\infty}^\infty
d\tau\ e^{-i(E_j - E_i) \tau} G(\cosh \tau).
\label{thermal}
\end{equation}
The first hint that~(\ref{thermal}) has something to do with
a thermal response is that the function $G$ is
periodic in imaginary time under $\tau \to \tau + 2 \pi i$, and
Green functions which are periodic in imaginary time
are thermal Green functions.

To investigate the nature of a thermal state, let us
suppose it were true (as will be demonstrated shortly) that
\begin{equation}
\dot{P}(E_i \to E_j) = \dot{P}(E_j \to E_i) e^{-\beta(E_j - E_i)},
\label{probij}
\end{equation}
and that the energy levels of the detector were thermally populated, so
that
\begin{equation}
N_i=N e^{-\beta E_i},
\label{thermpop}
\end{equation}
where $N$ is some normalization factor.
Then it is clear that the total  transition rate $R$
from $E_i$ to $E_j$
is the same as from $E_j$ to $E_i$:
\begin{equation}
R(E_i\to E_j)=N e^{-\beta E_i} \dot{P}(E_i\to E_j)=
R(E_j\to E_i),
\label{detail}
\end{equation}
which is the principle of detailed balance in a thermal ensemble.
In other words, if the transition probabilities are related
by~(\ref{probij}) and the population of the states is thermal as
in~(\ref{thermpop}), then there is no change in the probability distribution
for the energy levels with time.  So~(\ref{probij}) describes the transition
probabilities  of a system in a thermal bath of particles at
temperature $T = 1/\beta$.

\begin{figure}[hbtp]
    \centering
    \includegraphics[width=3.0truein]{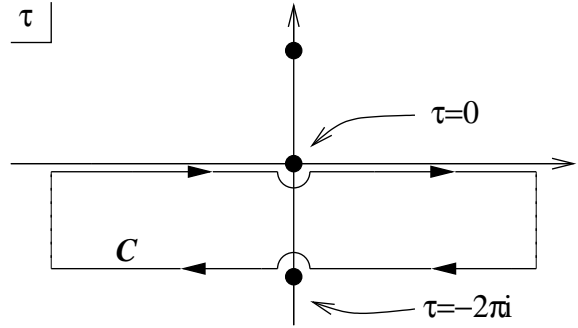}
    \caption[]{The integrand in~(\ref{thermal}) has
    singularities in the complex $\tau$-plane at $\tau = 2 \pi i n$
    for any integer $n$.  This figure shows the contour $C$
    used in the integral~(\ref{integraltwo}).  The dotted lines
    signify the closure of the contour at infinity.}
    \label{contourfig}
\end{figure}

Let us now show that~(\ref{probij}) holds for the transition
probabilities calculated in~(\ref{thermal}).
The integrand in~(\ref{thermal}) has singularities in the complex
$\tau$-plane at $\tau = 2 \pi i n$ for any integer $n$.
Consider integrating the function
$e^{-i(E_j - E_i) \tau} G(\cosh \tau)$
around the contour shown in figure \ref{contourfig}.

Since the total integral around this contour is zero,
we have
\begin{equation}
\int_{-\infty}^{\infty} d\tau\ e^{-i(E_j-E_i) \tau} G(\cosh \tau)+
\int_{+\infty-i \beta}^{-\infty-i\beta}
d\tau\ e^{-i(E_j-E_i) \tau} G(\cosh \tau)=0,
\label{integraltwo}
\end{equation}
where $\beta = 1/2 \pi$. The contour in figure \ref{contourfig}
corresponds to the pole prescription for the Wightman function as discussed
in section \ref{vacua}.
Now redefining the variable of integration in the second
integral as $\tau'=-\tau-i \beta$
we get precisely the desired relation~(\ref{probij}).

Although we performed this calculation only for an observer
stationary at the south pole, all timelike geodesics in de Sitter
space are related to each other by the SO$(d,1)$ de Sitter
isometry group.
Since the Green function used in this calculation is
de Sitter invariant, the result for the temperature is the same
for any observer moving along a timelike geodesic.
We conclude that any geodesic observer in de Sitter space will feel
that she/he is in a thermal bath of particles at
a temperature
\begin{equation}
T_{dS}=\frac{1}{2 \pi \ell},
\label{dstem}
\end{equation}
where we have restored the factor of the de Sitter radius $\ell$ by
dimensional analysis.

\subsection{Entropy}
\label{entropy}

In this subsection we will associate an entropy to de Sitter space. 
We will restrict our attention to \dst, where
the analysis simplifies considerably.

For the case of black holes
one can use similar methods as those in the previous section
to calculate the
temperature $T_{\rm BH}$ of the black hole. The black hole entropy
$S_{\rm BH}$
can then be found by integrating the thermodynamic relation
\begin{equation}
\frac{dS_{\rm BH}}{dE_{\rm BH}}=\frac{1}{T_{\rm BH}},
\label{dsde}
\end{equation}
where $E_{\rm BH}$ is the energy or mass of the black hole.
So if you know the value of the temperature just for one value of $E_{\rm BH}$
you will not be able to get the entropy, but if you know it as a function
of the black hole mass then you can simply integrate~(\ref{dsde})
to find the entropy.  The constant of integration is determined by
requiring that a black hole of zero mass has zero entropy.

So for de Sitter space one would expect to use the relation
\begin{equation}
\frac{dS_{\rm dS}}{dE_{\rm dS}} = \frac{1}{T_{\rm dS}}
\label{dsds}
\end{equation}
to find the entropy $S_{\rm dS}$.
The problem in de Sitter space is that
once the coupling constant of the theory is chosen there is
just one de Sitter
solution, whereas in the black hole case there is a whole one
parameter family of solutions labeled by the mass of the black hole,
for fixed coupling constant.
In other words, what is $E_{\rm dS}$ in~(\ref{dsds})?
One might try to vary the cosmological constant,
but that is rather unphysical as it is the coupling constant.
One would be going from one theory
to another instead of from one configuration in the theory
to another configuration in the same theory.

Let us instead follow Gibbons and Hawking \cite{Gibbons:1977mu} and
use the one parameter family of Schwarzschild-de Sitter solutions
to see how the temperature varies as a function of the parameter
$E$ labeling the mass of the black hole.

{\it Exercise~3.}
The SdS${}_3$ solution in static coordinates is
\begin{equation}
ds^2=-(1-8GE-r^2) dt^2+\frac{dr^2}{(1-8GE-r^2)}+r^2 d\phi^2.
\end{equation}
Find a Green function for SdS${}_3$
by analytic continuation from the smooth Euclidean solution.
Show that this Green function
is periodic in imaginary time with periodicity
\begin{equation}
\tau \to \tau+\frac{2 \pi i}{\sqrt{1-8GE}}.
\end{equation}

 From the exercise and the discussion in the previous section we conclude
that the temperature associated with the Schwarzschild-de Sitter
solution is
\begin{equation}
T_{\rm SdS}=\frac{\sqrt{1-8GE}}{2 \pi}.
\end{equation}
Using the formula
\begin{equation}
\frac{dS_{\rm SdS}}{dE}=\frac{1}{T_{\rm SdS}},
\label{wrongsign}
\end{equation}
and writing the result in terms of the area $A_H$ of the de Sitter 
horizon at $r_H=\sqrt{1-8GE}$ which is given by
\begin{equation}
\sqrt{1-8 G E}=\frac{A_H}{2 \pi},
\end{equation}
one finds that the entropy is equal to
\begin{equation}
S_{\rm SdS} =-\frac{A_H}{4G}.
\label{wrongs}
\end{equation}
This differs by a minus
sign from the famous formula~(\ref{bh})!
What did we do wrong?  Gibbons and Hawking suggested that to
get the de Sitter entropy we should use not~(\ref{wrongsign})
but instead
\begin{equation}
\label{right}
\frac{dS_{\rm SdS}}{d(-E_{\rm dS})}
=\frac{1}{T_{\rm SdS}}.
\end{equation}
This looks funny but
in fact there is a very good reason for using this new formula.

\begin{figure}[hbtp]
    \centering
    \includegraphics[width=1.6truein]{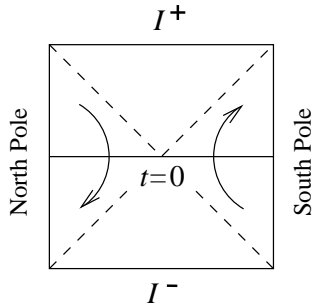}
    \caption[]{The energy associated to 
the Killing vector $\partial/\partial t$
    (indicated
    by the arrows) along the spacelike slice $t=0$ (solid
    line) must vanish.  If we ascribe positive energy to a
    positive deficit angle at the south pole, then we must
    ascribe negative energy to a positive deficit angle at
    the south pole since the Killing vector $\partial/\partial t$
    runs
    in the opposite direction behind the horizon.}
    \label{ddtsign}
\end{figure}

The de Sitter entropy, although we don't know exactly how to think
about it, is supposed to correspond to the entropy of the
stuff behind the horizon which we can't observe.
Now in general relativity
the expression for the
energy on a surface is the integral of a total derivative, which reduces
to a surface integral on the boundary of the surface, and hence vanishes on
any closed surface.
Consider a closed surface in de Sitter space such as the one shown
in figure \ref{ddtsign}.
If we put something with positive energy on the south
pole, then necessarily there will be some negative energy on the
north pole.

This can be seen quite explicitly in the Schwarzschild-de Sitter solution.
With no black hole, the spacelike slice in figure \ref{ddtsign}
is an $S^2$,
but we saw in one of the exercises that in the SdS${}_3$ solution
there is a positive deficit angle at both the north and south poles.
If we ascribe positive energy to the positive deficit angle at the
south pole, then because the Killing vector $\partial / \partial t$ used
to define the energy changes direction across the horizon, we are
forced to ascribe negative energy to the positive deficit angle
at the north pole.

Therefore the northern singularity of Schwarzschild-de Sitter
behind the horizon actually carries negative energy.
In~(\ref{wrongsign}) we varied with respect to the energy at the
south pole, and ended up with the wrong sign in~(\ref{wrongs}), but
if we more sensibly vary with respect to the energy at the north pole,
then we should use the formula~(\ref{right}).
Then we arrive at the entropy for Schwarzschild-de Sitter
\begin{equation}
S_{\rm SdS} = \frac{A_H}{4G} = \frac{\pi}{2 G} \sqrt{1 - 8 G E}.
\end{equation}
The integration constant has been chosen so that the entropy vanishes
for the maximal energy $E={1 \over 8G}$ at which value the
deficit angle is $2\pi$ and the space has closed up.

In conclusion we see that the area-entropy law~(\ref{bh}) indeed applies to
three dimensional Schwarzschild-de Sitter.

\section{Quantum Gravity in De Sitter Space}
\label{secqg}

So far we have discussed well established and understood results
about classical de Sitter space and quantum field theory in a
fixed de Sitter background. Now we turn to the more challenging
problem of quantum gravity in de Sitter space, about which little
is established or understood.

In this section we will give a pedagogical discussion of several aspects 
of some recent 
efforts in this direction \cite{dscft} (some similar ideas appeared in
\cite{Antoniadis:1997fu,Hull:1998vg,Bousso:1999cb,Banados:1999tb,
wit,Mazur:2001aa}).
We will argue that quantum gravity in
\dst\ can be described by a two dimensional conformal field
theory, in the sense that correlation functions of an operator
$\phi$ inserted at points $x_i$ on $\scri^-$ or $\scri^+$ are
generated by a two dimensional Euclidean CFT:
\begin{equation}
\langle \phi(x_1) \cdots \phi(x_i)\rangle_{{\rm dS}_3} \leftrightarrow
\langle {\cal{O}}_\phi(x_1) \cdots {\cal{O}}_\phi(x_i) \rangle_{S^2},
\label{wht}
\end{equation}
where ${\cal{O}}_\phi$ is an operator in the CFT associated to the
field $\phi$.
Equation~(\ref{wht}) expresses
the dS/CFT correspondence. 
The tool which will allow us to reach this
conclusion is an analysis of the asymptotic symmetry group for
gravity in \dst. Parallel results pertain in arbitrary dimension,
but the three dimensional case is the richest because of the
infinite dimensional nature of the $\scri^\pm$  conformal group.
The results  of this section are largely contained in
\cite{dscft}
except for the derivation of the asymptotic boundary
conditions for \dst , which were assumed/guessed without
derivation in \cite{dscft}.

\subsection{Asymptotic Symmetries}

Consider a simple U$(1)$ gauge theory in flat Minkowski space.
A gauge transformation which goes to zero
at spatial infinity will annihilate physical states (this
is just the statement that a physical state is gauge invariant), while a
gauge transformation which goes to a constant at spatial infinity
will act nontrivially on the states.  In fact the generator will be
proportional
to the charge operator, by Noether's theorem.

It is useful therefore to consider the so-called asymptotic symmetry
group (ASG), which is defined as
the set of allowed symmetry transformations modulo
the set of trivial symmetry transformations.
\begin{equation}{\rm ASG}= \frac{\rm Allowed~~Symmetry~~Transformations}
{\rm Trivial~~Symmetry~~Transformations}.
\end{equation}Here `allowed' means
that the transformation is consistent with the boundary conditions
that we have specified for the fields in the theory, and `trivial'
means that the generator of the transformation vanishes after we
have implemented the constraints---for example asymptotically
vanishing gauge transformations in the example of the previous
paragraph. The states and correlators of the theory clearly must lie in
representations of the ASG. Of course one must know the details of
the theory to know which representations of the ASG actually
appear, but in some cases a knowledge of the ASG already places
strong constraints on the theory.

In this section we will see that the ASG of quantum gravity in
\dst\ is the Euclidean conformal group in two dimensions. Since
this group acts on $\scri^\pm$, this means that correlators with
points on $\scri^\pm$ are those of a conformal field theory, and 
the correspondence (\ref{wht}) is simply an expression of diffeomorphism
invariance of the theory. 
Although we will not learn anything about the details of this
theory, the fact that the conformal group in two dimensions is
infinite dimensional already strongly constrains the physics.

In quantum gravity
the relevant gauge symmetry is diffeomorphism invariance, and in
de Sitter space the only asymptotia are $\scri^\pm$.
Therefore we need to consider diffeomorphisms in \dst\
which preserve the boundary conditions on the metric at $\scri^\pm$
but do not fall off so fast that they act trivially on physical
states.  The analogous problem for three dimensional anti-de Sitter
space was solved long ago by Brown and Henneaux \cite{bh}.
The result for de Sitter differs only by a few signs. However the
physical interpretation in the \dst\ case is very different
from that of AdS${}_3$, and remains to be fully understood.

\subsection{De Sitter Boundary Conditions 
and the Conformal Group}

Our first task is to specify the boundary conditions appropriate
for an asymptotically \dst\ spacetime. In general specification of
the boundary conditions is part of the definition of the theory,
and in principle there could be more than one choice. However if
the boundary conditions are too restrictive, the theory will
become trivial. For example in 4d gravity, one might try to demand
that the metric fall off spatially as ${1 \over r^2}$. This would
allow only zero energy configurations and hence the theory would
be trivial. On the other hand one might try to demand that it fall
off as ${1 \over \sqrt{r}}$. Then the energy and other symmetry
generators are in general divergent, and it is unlikely any sense can
be made of the theory. So the idea is to make the falloff as weak
as possible while still maintaining finiteness of the generators.

Hence we need to understand the surface integrals which generate
the diffeomorphisms of \dst. A convenient and elegant formalism
for this purpose was developed by Brown and 
York \cite{by,Brown:2000dz}
(and applied to AdS$_3$ in
\cite{Balasubramanian:1999re,deHaro:2001xn}). They
showed that bulk diffeomorphisms are generated by appropriate
moments of a certain stress tensor which lives on the boundary of
the spacetime.\footnote{Brown
and York mainly consider a timelike boundary, but their results can be
extended to the spacelike case.} We will define an asymptotically
\dst\ spacetime to
be one for which the associated stress tensor, and hence all the
symmetry generators, are finite.

The Brown-York stress tensor for \dst\ with $\ell = 1$ is given
by\footnote{We caution
the reader that the generalization of~(\ref{brownyork})
to $d>3$ or to theories with matter is not entirely straightforward
\cite{deHaro:2001xn}.}
\begin{equation}
T_{\mu \nu} = {1 \over 4 G} \left[ K_{\mu \nu}
- (K + 1) \gamma_{\mu\nu}\right].
\label{brownyork}
\end{equation}
Here $\gamma$ is the induced metric on the boundary $\scri^-$ and
$K$ is the trace of the extrinsic curvature $K_{\mu \nu} = -
\nabla_{(\mu} n_{\nu)} = - {1 \over 2} {\cal L}_n \gamma_{\mu\nu}$
with $n^\mu$ the outward-pointing unit normal. (\ref{brownyork}) vanishes
identically for vacuum \dst\ in planar
coordinates~\begin{equation} \label{dsmet} ds^2=-dt^2
+e^{-2t}dzd\bar z.
\end{equation} For a perturbed metric $g_{\mu \nu} +
h_{\mu \nu}$ we obtain the Brown-York stress tensor
\begin{eqnarray}
T_{z z} &=& {1 \over
4 G} \left[h_{zz} - \partial_z h_{tz} + {1 \over 2} \partial_t h_{zz}\right] +
{\cal{O}}(h^2),\cr
T_{z \bar{z}} &=& {1 \over 4 G}
\left[ {1 \over 4} e^{-2t} h_{tt} - h_{z \bar{z}} +
{1 \over 2}(
\partial_{\bar{z}} h_{tz} +
\partial_z h_{t \bar{z}}- \partial_t h_{z
\bar{z}})\right] + {\cal{O}}(h^2).
\label{byresult}
\end{eqnarray}
Details of this calculation are given in appendix A.
Requiring the stress tensor to be finite evidently leads to the
boundary conditions
\begin{eqnarray}
g_{z \bar z}&=&{e^{-2t} \over 2} +{\cal O}(1),\cr
g_{tt}&=&-1 +{\cal O}( e^{2t} ),\cr
g_{zz}&=&{\cal O}( 1 ),\cr
g_{tz}&=&{\cal O}(1 ).
\label{boundaryconditions}
\end{eqnarray}
It is not hard to see that the most general diffeomorphism
$\zeta$ which preserves the boundary conditions~(\ref{boundaryconditions})
may be written as
\begin{equation}
\zeta = U \partial_z + {1 \over 2} U' \partial_t +
{\cal{O}}(e^{2 t}) +
{\rm complex~conjugate},
\label{diff}
\end{equation}
where $U = U(z)$ is holomorphic in $ z$.\footnote{We allow isolated 
poles in $z$. In principle this should be  carefully 
justified (as (\ref{boundaryconditions}) is violated very near the 
singularity), and we have not done so here. A parallel issue arises in 
AdS$_3$/CFT$_2$.}  A diffeomorphism of
the form~(\ref{diff}) acts on the Brown-York stress tensor as
\begin{equation}
\delta_\zeta T_{zz} = -U \partial T_{zz} -2 U' T_{zz} - {1 \over
8G} U'''. \label{deltat}
\end{equation}
The first two terms are those appropriate for an operator of
scaling dimension two. The third term is the familiar linearization
of the anomalous Schwarzian derivative term corresponding to a
central charge
\begin{equation}
c=\frac{3 l}{2 G},
\end{equation}
where we have restored the power of $\ell$.\footnote{Parallel derivations
of 
the central charge for AdS were given in
\cite{Henningson:1998gx,Balasubramanian:1999re}.} 
Note that the
${\cal{O}}(e^{2 t})$ terms in~(\ref{diff}) do not contribute
in~(\ref{deltat}). Therefore they are trivial diffeomorphisms, in
the sense described above. We conclude that the asymptotic
symmetry group of \dst\ as generated by~(\ref{diff}) is the
conformal group of the Euclidean plane.

The last boundary condition~(\ref{boundaryconditions}) differs
from the condition $g_{tz} = {\cal{O}}(e^{2 t})$ assumed in
\cite{dscft} and  obtained by analytically continuing the
AdS${}_3$ boundary conditions of Brown and Henneaux \cite{bh} from
anti-de Sitter to de Sitter space. The resolution of this apparent
discrepancy comes from noting that if $g_{tz} \to f$ on the
boundary
 where $f =f(z,\bar{z})$
is an arbitrary function, then applying the diffeomorphism $\zeta
= e^{2 t} f \partial_{\bar{z}}$ gives $\delta_\zeta g_{tz} =
{\cal{O}}(e^{2t})$.  Therefore one can always set the component
$g_{tz}$ of the metric to be ${\cal{O}}(e^{2t})$ with a trivial
diffeomorphism. In other words, if $g_{tz} = {\cal{O}}(1)$, then
in fact one can always choose a gauge in which $g_{tz} =
{\cal{O}}(e^{2t})$. Exploiting this freedom one can impose the
asymptotic boundary conditions
\begin{eqnarray}
g_{z \bar z}&=&{e^{-2t} \over 2} +{\cal O}(1),\cr g_{tt}&=&-1
+{\cal O}( e^{2t} ),\cr g_{zz}&=&{\cal O}( 1 ),\cr g_{tz}&=&{\cal
O}(e^{2t} ), \label{daryconditions}
\end{eqnarray}
as given in \cite{dscft}.

A special case of~(\ref{diff}) is the choice
\begin{equation}
U=\alpha +\beta  z +\gamma z^2,
\label{drfv}
\end{equation}
where $\alpha ,~\beta  ,~\gamma $ are complex constants.
In this case $U^{\prime\prime\prime}$ vanishes, and the \dst\ metric
is therefore invariant.    These transformations generate the
SL$(2,{\bf{C}})$ global isometries of \dst.

Where do conformal transformations come from? Recall that a
conformal transformation in two dimensions is a combination of an
ordinary diffeomorphism and a Weyl transformation. In two
dimensions a diffeomorphism acts as
\begin{equation}
g_{z \bar{z}} \to \frac{d z'}{dz} \frac{d \bar{z}'}{d
\bar{z}} g_{z' \bar{z}'},
\label{diffeo}
\end{equation}
and a Weyl transformation acts as
\begin{equation}
g_{z \bar{z}} \to e^{2 \phi} g_{z \bar{z}}.
\label{weyl}
\end{equation}
A conformal transformation is just an ordinary
diffeomorphism~(\ref{diffeo}) followed by a Weyl
transformation~(\ref{weyl})
with $\phi$ chosen so that $g_{z \bar{z}} \to g_{z' \bar{z}'}$
under the combined transformation.

Now if we look at what the diffeomorphism $\zeta$ defined
in~(\ref{diff}) does, we see that the first term $U \partial_z$
generates a holomorphic diffeomorphism of the plane. Now the form
of the metric~(\ref{dsmet}) makes it clear that this can be
compensated by a shift in $t$, which from the point of view of the
$z$-plane is a Weyl transformation. This accounts for the second
term $\frac{1}{2} U' \partial_t$ in $\zeta$. So a diffeomorphism
in \dst\ splits into a tangential piece, which acts like an
ordinary diffeomorphism of the complex plane, and a normal piece,
which acts like a Weyl transformation. A three dimensional
diffeomorphism is thereby equivalent to a two dimensional conformal
transformation.  

Since $U(z)$ was arbitrary, we conclude that the asymptotic
symmetry group of gravity in \dst\ is the conformal group of the
complex plane.  The isometry group is the SL$(2,{\bf C})$ subgroup
of the asymptotic symmetry group. In particular, the ASG is
infinite dimensional, a fact which highly constrains quantum
gravity on \dst.  This is particular to the three dimensional
case, since in higher dimensional de Sitter space the ASG is the
same as the isometry group SO$(d,1)$.

We conclude these lectures with a last

{\it Exercise~4}.

(a)  Find an example of string theory on de Sitter space.

 (b)
Find the dual conformal field theory.

\ifpreprint
\vskip 0.25in
\noindent
{\bf Acknowledgments.}
We would like to thank C. Bachas, A. Bilal, F. David, M. Douglas,
    and N. Nekrasov for organizing a very pleasant and productive
    summer school and for arranging financial support.
It is a pleasure to thank R. Bousso and A. Maloney for useful discussions,
and we are also grateful to C. Herzog, I. Low,
L. McAllister and I. Savonije for comments on the
manuscript.
This work was supported in part by DOE grant
DE-FG02-91ER40655.
M.S. is also supported by
    DOE grant DE-FG02-91ER40671, and
    A.V. is also supported by INTAS-OPEN-97-1312.
\fi

\appendix

\section{Calculation of the Brown-York Stress Tensor}

We wish to calculate the Brown-York stress
tensor~(\ref{brownyork}) for a metric
which is a small perturbation of \dst.
We write the metric in planar coordinates $(t,z,\bar{z})$ as
\begin{equation}
\label{eq:gplush}
ds^2 = g_{\mu\nu} dx^\mu dx^\nu = - dt^2 + e^{-2t} dz d\bar{z}
+ h_{\mu \nu} dx^\mu dx^\nu,
\end{equation}
and we will always drop terms of order ${\cal{O}}(h^2)$.
We can put~(\ref{eq:gplush}) into the form
\begin{equation}
ds^2 = - N^2 dt^2 + \gamma_{ij} (dx^i + N^i dt)(dx^j + N^j dt),
\end{equation}
where the lapse and shift functions are given by
\begin{equation}
N = 1 - {1 \over 2} h_{tt}, ~~~~~
N^z = 2 e^{2 t} h_{t \bar{z}}, ~~~~~
N^{\bar{z}} = 2 e^{2 t} h_{t z},
\end{equation}
and the induced metric on the boundary $\scri^-$ is
\begin{equation}
\label{eq:gammaij}
\gamma_{zz} = h_{zz}, ~~~~~ \gamma_{z \bar{z}} = {1 \over 2}
e^{-2 t} + h_{z \bar{z}},
~~~~~ \gamma_{\bar{z} \bar{z}} = h_{\bar{z} \bar{z}}.
\end{equation}
The outward pointing unit normal vector to the boundary is
\begin{equation}
n^\mu = {1 \over N}\left(-1, N^z, N^{\bar{z}}\right)
= \left( -1 - {1 \over 2} h_{tt}, 2 e^{2 t} h_{t \bar{z}},
2 e^{2 t} h_{tz} \right).
\end{equation}
Upon lowering the indices, we have
\begin{equation}
n_\mu = \left( 1 - {1 \over 2} h_{tt}, 0, 0\right)
\end{equation}
and we  use the formula $K_{\mu\nu}
= - {1 \over 2} (\nabla_\mu n_\mu
+ \nabla_\nu n_\mu)$ to obtain
\begin{eqnarray}
\label{bykij}
K_{zz} &=& - \partial_z h_{tz} + {1 \over 2}
\partial_t h_{zz},\cr
K_{z \bar{z}} &=& - {1 \over 2} e^{-2t} (1 + {1 \over 2} h_{tt})
- {1 \over 2}
\left(\partial_{\bar{z}}
h_{tz} + \partial_z h_{t \bar{z}} - \partial_t h_{z \bar{z}}
\right).
\end{eqnarray}
The trace is
\begin{equation}
\label{byk}
K = g^{\mu\nu} K_{\mu\nu}
= \gamma^{ij} K_{ij} = -2 - h_{tt} +
4 e^{2 t} h_{z \bar{z}} - 2 e^{2 t} \left(
 \partial_{\bar{z}} h_{tz} + \partial_z h_{t \bar{z}} -
\partial_t h_{z \bar{z}}\right).
\end{equation}
Plugging~(\ref{bykij}) and~(\ref{byk}) into~(\ref{brownyork})
gives the desired result~(\ref{byresult}).

\end{document}